\documentclass[5p,numbers,sort&compress,preprint]{elsarticle}
\usepackage{graphicx,subfigure,amsbsy,amsmath,amssymb,color}
\begin{document}

\title{Insights on the cuprate high energy anomaly observed in ARPES}
%\title{Doping dependence of the cuprate high energy anomaly observed in ARPES}
%\title{Role of strong correlations in the cuprate high energy anomaly}

\author[SIMES,UND]{B.~Moritz\corref{cor1}\fnref{email}}
\author[SIMES,Waterloo]{S.~Johnston}
\author[SIMES,Geballe]{T.~P.~Devereaux}

\cortext[cor1]{Corresponding author}  
\fntext[email]{E-mail address:  moritzb@slac.stanford.edu}

\address[SIMES]{Stanford Institute for Materials and Energy Science, SLAC National Accelerator Laboratory, Menlo Park, CA 94025, USA}
\address[UND]{Department of Physics and Astrophysics, University of North Dakota, Grand Forks, ND 58202, USA}
\address[Waterloo]{Department of Physics and Astronomy, University of Waterloo, Waterloo, ON N2L 3G1, Canada}
\address[Geballe]{Geballe Laboratory for Advanced Materials, Stanford University, Stanford, CA 94305, USA} 

\date{\today}

\begin{abstract}
Recently, angle-resolved photoemission spectroscopy has been used to highlight an anomalously large band renormalization at high binding energies in 
cuprate superconductors: the high energy ``waterfall" or high energy anomaly (HEA).  The anomaly is present for both hole- and electron-doped 
cuprates as well as the half-filled parent insulators with different energy scales arising on either side of the phase diagram. While photoemission 
matrix elements clearly play a role in changing the aesthetic appearance of the band dispersion, \emph{i.e.}~creating a ``waterfall"-like appearance, 
they provide an inadequate description for the physics that underlies the strong band renormalization giving rise to the HEA.  Model calculations of 
the single-band Hubbard Hamiltonian showcase the role played by correlations in the formation of the HEA and uncover significant differences in the 
HEA energy scale for hole- and electron-doped cuprates.  In addition, this approach properly captures the transfer of spectral weight accompanying 
doping in a correlated material and provides a unifying description of the HEA across both sides of the cuprate phase diagram.  We find that the 
anomaly demarcates a transition, or cross-over, from a quasiparticle band at low binding energies near the Fermi level to valence bands at higher 
binding energy, assumed to be of strong oxygen character.  
\end{abstract}

\begin{keyword}
ARPES \sep quantum Monte Carlo \sep Hubbard \sep strong correlations 
\PACS 74.72.-h \sep 79.60.-i \sep 74.25.Jb \sep 71.10.Fd
\end{keyword}

\maketitle

Advancements in angle-resolved photoemission spectroscopy, an important probe of electronic structure, \cite{ARPES_Rev} have impacted significantly 
the study of strongly correlated materials.  High resolution experiments, at binding energies up to $1$ eV and higher, made possible by these 
advances, have revealed the presence of a ``waterfall"-like structure with a characteristic kink at intermediate binding energies -- the high energy 
anomaly (HEA) -- in the dispersion of high $T_{c}$ 
superconductors.~\cite{Non_anomaly,Graf_anomaly,Feng_anomaly,Chang_anomaly,Valla_anomaly,Borisenko_anomaly,Zhang_anomaly,Pan_anomaly,CKim_anomaly,Moritz_anomaly,Fujimori_anomaly}  
Found universally in hole-doped compounds, the characteristic HEA energy scale is $\sim 300$ meV with similar a dispersion anomaly observed in the 
half-filled parent insulators.~\cite{Ronning_anomaly}  In contrast to earlier low energy studies focusing on dispersion kinks under $\sim 100$ meV, 
\cite{ARPES_Zhou} interpreted as signatures of coupling to low energy bosonic modes, \cite{Devereaux_kink,Valla_kink,Norman_kink} the extrapolated 
band bottom has a value larger than that obtained from band structure calculations \cite{Non_anomaly,Graf_anomaly} and the energy scale associated 
with the anomaly would tend to rule-out coupling to similar bosonic modes as the origin of the HEA.  

More recently, anomalies have been found in electron-doped compounds at approximately twice the energy scale. 
\cite{Pan_anomaly,CKim_anomaly,Moritz_anomaly,Fujimori_anomaly}  Taken together with results from hole-doped and half-filled parent materials, these 
findings raise questions about the origin or mechanism of the HEA, given the apparent dichotomy in energy scales between electron- and hole-doped 
compounds, and what role, if any, many body effects may play.  Focusing primarily on the HEA in hole-doped compounds, a number of theories have been 
advanced including spin-charge separation, \cite{Graf_anomaly} in-gap band-tails, \cite{Alexandrov_HEA} spin polarons, \cite{Manousakis_HEA} coupling 
to spin fluctuations, \cite{Valla_anomaly,QMC,Paramagnon_anomaly} phonons, \cite{Feng_anomaly} or plasmons, \cite{Bansil_plasmon} strong correlation 
or ``Mott" physics, \cite{Non_anomaly,DMFT,Tan_HEA,Lanczos,LDA+DMFT} and even extrinsic effects associated with photoemission matrix 
elements.~\cite{Borisenko_anomaly}  

Photoemission matrix elements clearly play a role in modifying the appearance of the HEA, complicating the analysis.  The kink- or ``waterfall"-like 
structure found in the first Brillouin zone (BZ) in experiments performed using synchrotron sources appears instead in newer experiments performed 
using low photon energy, laser sources as a band ``break-up" or cross-over between a shallow, dispersing band crossing the Fermi level and a higher 
binding energy feature near the $\Gamma$-point.~\cite{Zhang_anomaly} Matrix element effects also influence the results of investigations in higher 
BZs which find a shallow band with a characteristic ``Y" appearance near the zone center, rather than a true ``waterfall".~\cite{Borisenko_anomaly}  
However, the shallow, dispersing band near the Fermi level, together with the cross-over at the HEA energy scale seen in both hole- and 
electron-doped cuprates, \cite{Zhang_anomaly,Moritz_anomaly,Fujimori_anomaly} originates from intrinsic band renormalization effects and not 
extrinsic mechanisms that merely serve to change the appearance of features with changing experimental conditions.~\cite{Bansil_ME_effect}   Those 
theoretical scenarios based on weak coupling to high energy bosonic modes 
\cite{Feng_anomaly,Valla_anomaly,Bansil_plasmon,QMC,Paramagnon_anomaly,Bansil_ME_effect} have an appeal based on the kink-like appearance of the HEA, 
recalling earlier efforts aimed at explaining the origin of the low-energy kink in cuprates. \cite{Devereaux_kink,ARPES_Zhou,Valla_kink,Norman_kink}  
However, coupling to these modes, such as spin fluctuations, which would generally satisfy the energy scale for the HEA in hole-doped compounds, 
fails to account for the dichotomy in energy scales between hole- and electron-doped materials.  However, spin fluctuations should play 
a prominent role in the renormalization mechanism forming the shallow, dispersing quasiparticle band for either hole- or electron-doped systems.   

To investigate the influence that many-body effects, \emph{i.e.}~strong correlations, may have on the origin of the HEA we investigate the 
single-particle spectral function of the single-band Hubbard model, building upon and adding to the information obtained from much earlier 
investigations.~\cite{Hanke_1,Hanke_2,Dagotto_RMP}  Using quantum Monte Carlo \cite{DQMC_1,DQMC_2} and the maximum entropy method (MEM) for analytic 
continuation, \cite{Jarrell_Guber_MEM,Alex_MEM} we study the spectral function for various values of electron filling.  Our results indicate that the 
HEA can be connected to doping and the accompanying transfer of spectral weight into the lower or upper Hubbard band (LHB or UHB) of hole- or 
electron-doped systems, respectively.  Doping leads to the formation of a quasi-particle band (QPB) at energies near the Fermi level, here set equal 
to zero energy, and the HEA represents a cross-over from this band to the LHB, playing the role of valence bands in the cuprates that would have 
substantial oxygen character.  Correlations and spectral weight transfers in the model lead to a natural asymmetry in the HEA energy scale between 
hole and electron doping, in agreement with experiment. The results also show qualitative similarities between the momentum dependence of the HEA 
from model calculations and that found experimentally.

The single-band Hubbard Hamiltonian, an effective, low energy model for the cuprates derived from down-folding multi-orbital models explicitly 
incorporating copper and oxygen degrees of freedom, \cite{Zhang_Rice,Anderson} has the form
\[%\begin{equation}
H=-\sum_{ij,\sigma}t_{ij}c^{\dag}_{i,\sigma}c_{j,\sigma}-\mu\sum_{i,\sigma}n_{i,\sigma}+\sum_{i}U(n_{i,\uparrow}-\frac{1}{2})(n_{i,\downarrow}-\frac{1}{2}),
\]%\end{equation}
where $c^{\dag}_{i,\sigma} (c_{i,\sigma})$ creates (annihilates) an electron with spin $\sigma$ at site $i$, the site occupation $n_{i,\sigma} = 
c^{\dag}_{i,\sigma}c_{i,\sigma}$ for each spin species equals $0$ or $1$, $\{t_{ij}\}$ is a set of tight-binding coefficients parameterizing the 
electron kinetic energy (in this study we restrict ourselves to nearest-neighbor, $t$, and next-nearest-neighbor, $t'$, terms), the chemical 
potential $\mu$ controls the electron filling, and the Hubbard repulsion $U$ controls the strength of electron-electron correlations.  The model is 
studied using determinant quantum Monte Carlo, \cite{DQMC_1,DQMC_2} an auxiliary-field technique.  This method yields the finite temperature, 
imaginary time propagator $G_{ij}(\tau)$ on a finite-size cluster with periodic boundary conditions.  

Estimates for this propagator in imaginary time come from individual Markov chains of the Monte Carlo process.  This imaginary time propagator must 
be Wick rotated, or analytically continued, to real frequencies to extract the spectral function.  Performing a discrete Fourier transform and 
treating the ensemble $\{G_{K}(\tau)\}$ obtained from individual Markov chains as a statistical sampling of the finite temperature propagator, the 
MEM, \cite{Jarrell_Guber_MEM,Alex_MEM} based on Bayesian inference, is used to obtain the single-particle spectral function $A(\mathbf{K},\omega)$ on 
the appropriate discrete momentum grid.  

The finite-size, square clusters used in this study have linear dimension $N=8$ ($64$-site clusters).  The corresponding momentum space grid $\{K\}$ 
has a point spacing of $\pi/4$.  The imaginary time interval, partitioned into L ``slices" of size $\Delta\tau = \beta/L$, runs from 0 to $\beta = 
1/T$, the inverse temperature, and $t$ serves as the energy unit of the problem. In this study, $\beta = 3/t$ and $\Delta\tau = 1/16\,t$. Once 
$A(\mathbf{K},\omega)$ has been obtained on the discrete momentum grid, the single-particle self-energy $\Sigma(\mathbf{K},\omega)$ can be extracted 
using Dyson's equation and the bare bandstructure corresponding to the tight-binding, model parameters.  Assuming a weak momentum dependence to the 
self-energy, an interpolation routine provides the value of $\Sigma(\mathbf{k},\omega)$ at an arbitrary point $\mathbf{k}$ in the BZ and Dyson's 
equation can be employed to compute $A(\mathbf{k},\omega)$.  

\begin{figure}[t*]
\centering
\includegraphics[width=\columnwidth]{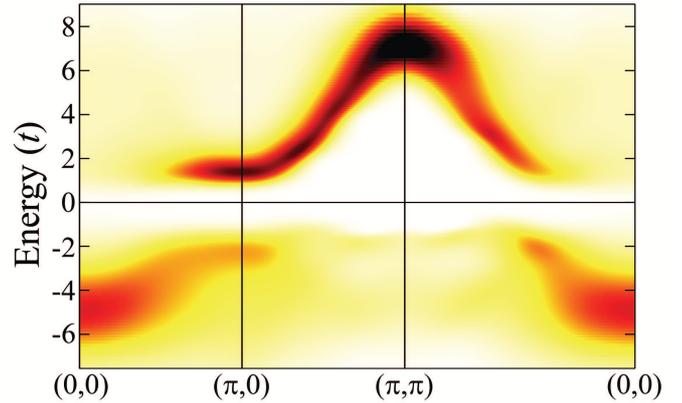}
\caption{\label{fig:figure1}Bandstructure of the undoped single-band Hubbard model along high symmetry directions in the BZ.  Model parameters: 
$t'=-0.30t$, $\mu=0.00t$, $U=8.00t$. The UHB has a pronounced dispersion along the $(\pi,0)-(\pi,\pi)$ and $(0,0)-(\pi,\pi)$ directions.  The LHB, 
broad and centered at the $\Gamma$-point, has similar dispersing features, better separated from the bulk of the LHB especially along the 
$(0,0)-(\pi,\pi)$ direction.  Note the well separated intensity near $(\pi/2,\pi/2)$ between $-1t$ and $-2t$ and that at $(\pi,0)$ near $-2t$.  These 
dispersing features are precursors to the QPB that forms upon electron or hole doping.  (Color online)}
\end{figure}

Figure \ref{fig:figure1} shows the bandstructure along high symmetry directions in the BZ for the undoped single-band Hubbard model with parameters 
$t'=-0.30t$, $\mu=0.00t$, and $U=8.00t$.  At the $\Gamma$-point and $(\pi,\pi)$ one finds the bulk of the incoherent LHB and UHB, respectively.  
Above the Fermi level, the UHB has dispersing branches along the $(\pi,0)-(\pi,\pi)$ and $(0,0)-(\pi,\pi)$ directions with significant spectral 
weight in the region near $(\pi,0)$.  Below the Fermi level, the bulk of the LHB spectral weight is concentrated near the $\Gamma$-point; however, 
there is less pronounced, but better separated, dispersing spectral weight along the $(0,0)-(\pi,0)$ and $(0,0)-(\pi,\pi)$ directions.  

There is a significant spectral weight in this dispersing feature near $(\pi/2,\pi/2)$ between $-1t$ and $-2t$ before losing intensity upon 
approaching the $\Gamma$-point and the bulk of the LHB.  A similar energy scale can be assigned to the dispersing feature near $(\pi,0)$. The 
behavior of the spectral intensity along the $(0,0)-(\pi,\pi)$ direction is qualitatively similar to that observed in experiment 
\cite{Ronning_anomaly} with a low binding energy feature near  $(\pi/2,\pi/2)$ crossing-over to the valence band at the $\Gamma$-point with the 
transition marked by the appearance of a ``waterfall" at intermediate binding energies.  These dispersing features are precursors to the QPB that 
appears upon electron or hole doping.

\begin{figure}[t*]
\centering
\includegraphics[width=\columnwidth]{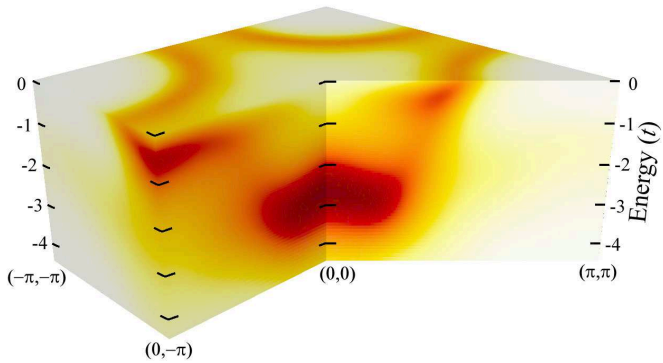}
\caption{\label{fig:figure2}Bandstucture of the hole-doped ($\sim14\%$) single-band Hubbard model along high symmetry directions in the BZ below the 
Fermi level.  Each face represents a different symmetry direction arranged in the proper geometric relationship to one another.  The top face shows 
the Fermi surface with a portion removed, exposing the bandstructure along the BZ axis (the $(0,0)-(0,-\pi)$ direction) and BZ diagonal (the 
$(0,0)-(\pi,\pi)$ direction).  Model parameters:  $t'=-0.30t$, $\mu=-2.50t$, $U=8.00t$.  In this figure, the spectral function $A(\mathbf{k},\omega)$ 
has been multiplied by the Fermi distribution function $f(\omega)$.  The dispersing QPB crosses the Fermi level and remains shallow, at low binding 
energies, giving way to the LHB, localized near the $\Gamma$-point, with a cross-over of weaker intensity at intermediate binding energies.  (Color 
online)}
\end{figure}

Figure \ref{fig:figure2} shows the bandstructure for the hole-doped single-band Hubbard model below the Fermi level along high symmetry directions in 
the BZ, arranged with the appropriate geometrical relationship.  Here the spectral function has been multiplied by the Fermi distribution function 
$f(\omega)=(\exp(\beta\omega)+1)^{-1}$.  The incoherent LHB, effectively localized near the $\Gamma$-point, has a weak tail of intensity extending 
toward the points $(\pi,\pi)$ and $(0,-\pi)$.  The decrease in intensity within these tails approximately coincides with the location in momentum 
space identified with the HEA.  

Along the $(0,0)-(\pi,\pi)$ direction, the QPB disperses across the Fermi level at $\sim(\pi/2,\pi/2)$.  Near $(\pi/4,\pi/4)$ the spectral intensity 
drops and the HEA appears as an apparent cross-over from the QPB to the LHB at an energy $\sim -0.5t$ to $-0.75t$, the HEA energy scale.  Along the 
$(0,0)-(0,-\pi)$ direction, the QPB is nearly non-dispersive at an energy $\sim -0.5t$.  At approximately $(0,-\pi/2)$ the QPB crosses-over to the 
LHB with a weak tail of intensity at low binding energy extending toward the $\Gamma$-point.  While the spectral intensity decreases in the QPB 
approaching the $\Gamma$-point, there is significant coexistence of the LHB and QPB as a function of momentum along these high symmetry cuts.  The 
spectral function has a ``waterfall"-like appearance even without including the effect of photoemission matrix elements, which serve to enhance the 
appearance by further suppressing intensity nearing the $\Gamma$-point.~\cite{Moritz_anomaly}

\begin{figure}[t*]
\centering
\includegraphics[width=\columnwidth]{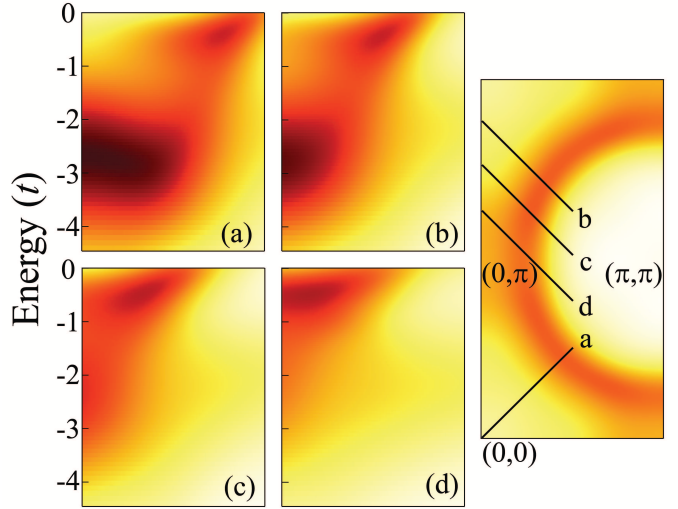}
\caption{\label{fig:figure3}Momentum dependence of the hole-doped spectral function multiplied by the Fermi distribution function along selected cuts 
in the BZ.  Model parameters:  $t'=-0.30t$, $\mu=-2.50t$, $U=8.00t$.  Panels (a)-(d) show the intensity (falsecolor scale) along the cuts highlighted 
in the right panel.  This sequence of cuts highlights the distinction between the QPB and the LHB and shows the evolution of the ``waterfall"-like 
appearance of the HEA moving along the BZ axis away from the $\Gamma$-point.  (Color online)}
\end{figure} 

The relationship between the QPB, the LHB, and the appearance of the HEA can be further explored studying the detailed momentum dependence of the 
spectral function.  Figure \ref{fig:figure3} shows the spectral function along selected momentum space cuts.  Cut (a) reproduces a part of the 
spectral function along the $(0,0)-(\pi,\pi)$ direction already encountered in Fig.~\ref{fig:figure2}.  Cuts (b)-(d) show the spectral function in 
momentum space, parallel to the $(0,0)-(\pi,\pi)$ direction, moving toward $(0,\pi)$.  While the energy scale of the LHB remains relatively 
unchanged, its spectral intensity progressively decreases.  The QPB, crossing the Fermi level, increasingly becomes easier to identify as a proper 
band with increased spectral weight approaching the BZ axis.  The ``waterfall"-like appearance of the HEA also evolves with changes to its momentum 
space position and progressive reduction in spectral intensity below the QPB.  Taken as a whole, the evolution of the QPB and HEA qualitatively agree 
with the results of experiment on hole-doped compounds, \cite{Non_anomaly,Graf_anomaly} including the evolution of spectral intensity and changes in 
momentum space position and robustness of the ``waterfall"-like appearance as a function of momentum.

\begin{figure}[t*]
\centering
\includegraphics[width=\columnwidth]{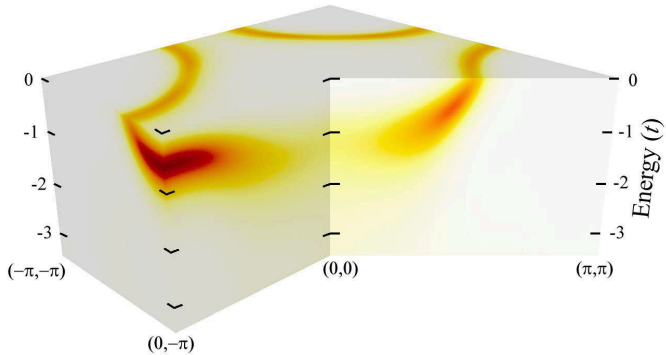}
\caption{\label{fig:figure4}Bandstucture of the electron-doped ($\sim17\%$) single-band Hubbard model along high symmetry directions in the BZ below 
the Fermi level.  The top face shows the Fermi surface with a portion removed.  Model parameters:  $t'=-0.25t$, $\mu=2.20t$, $U=8.00t$.  In this 
figure, the spectral function $A(\mathbf{k},\omega)$ has been multiplied by the Fermi distribution function $f(\omega)$.  The dispersing QPB crosses 
the Fermi level and remains shallow, at low binding energies approaching the $\Gamma$-point with diminishing intensity.  The LHB (not shown) at 
higher binding energies ($\sim -7.5t$) has a substantially reduced intensity due to spectral weight transfer into the QPB and UHB with electron 
doping.  The HEA energy scale is $\sim -1.0t$ to $-1.5t$, twice that found in the hole-doped system.  (Color online)}
\end{figure}

Comparison of calculation results for a hole-doped system to those for an electron-doped system reveal the dichotomy in HEA energy scales intrinsic 
to the single-band Hubbard model.  Figure \ref{fig:figure4} shows the spectral function multiplied by the Fermi distribution function for an 
electron-doped system with model parameters $t'=-0.25t$, $\mu=2.20t$, and $U=8.00t$.  The LHB (not shown) has a dramatically reduced intensity due to 
spectral weight transfers into the QPB and UHB upon electron doping.  This affects the ``waterfall"-like appearance of the HEA in the electron-doped 
calculation by reducing the spectral intensity at intermediate binding energies between the QPB and the LHB.  Previous results demonstrate that 
modifying the spectral intensity through the inclusion of approximate matrix elements leads to enhanced ``waterfall"-like character in the 
electron-doped dispersion.~\cite{Moritz_anomaly} For real-world experiments, the proper valence band, composed of significant oxygen character, 
should remain robust under electron-doping, increasing the likelihood of a prominent ``waterfall"-like appearance in the cross-over region at 
intermediate binding energies, in agreement with experiment.~\cite{Pan_anomaly,CKim_anomaly,Moritz_anomaly,Fujimori_anomaly}

The dispersive QPB shown in Fig.~\ref{fig:figure4} dips farther below the Fermi level than does its hole-doped counterpart, yielding an HEA energy 
scale  $\sim -1.0t$ to $-1.5t$ along the $(0,0)-(\pi,\pi)$ direction. Along the $(0,0)-(0,-\pi)$ direction, the spectral function appears to be quite 
similar to its hole-doped counterpart with an energy scale $\sim -0.5t$ near $(0,-\pi)$ with a weak dispersion in this region.  Together with 
decreasing intensity, the QPB along this direction disperses downward to and energy $\sim -1.5t$ at the $\Gamma$-point.  The chief contrast between 
the QPB in hole- and electron-doped systems, other than the energy scale, lies in their origin.  With hole doping, the chemical potential moves down 
into the LHB, or more precisely the dispersive features at lower binding energy, and the QPB originates from the dispersive precursors shown in 
Fig.~\ref{fig:figure1}; whereas in electron-doped systems, the chemical potential shifts in the opposite direction and the QPB arises from the 
precursors in the UHB.

\begin{figure}[t*]
\centering
\includegraphics[width=\columnwidth]{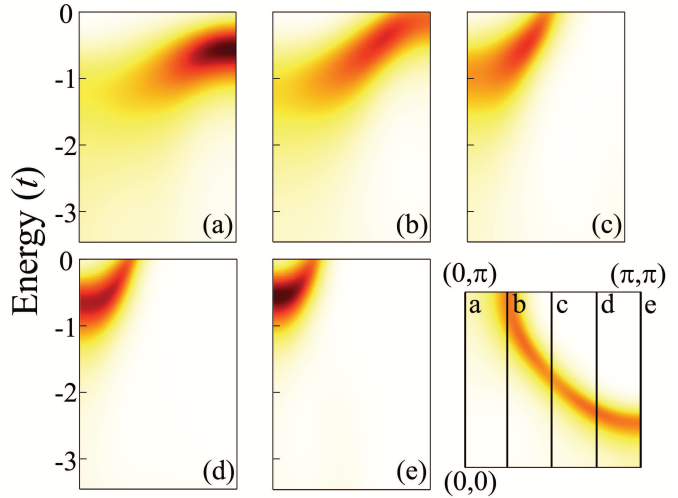}
\caption{\label{fig:figure5}Momentum dependence of the electron-doped spectral function multiplied by the Fermi distribution function along selected 
cuts in the BZ.  Model parameters:  $t'=-0.25t$, $\mu=2.20t$, $U=8.00t$.  Panels (a)-(e) show the intensity (falsecolor scale) along the cuts 
highlighted in the lower right panel.  The lack of significant spectral intensity in the LHB suppresses the ``waterfall"-like appearance of the HEA.  
(Color online)}
\end{figure}

Figure \ref{fig:figure5} shows the momentum dependence of the QPB along selected cuts in the BZ.  Like the hole-doped system, the spectral intensity 
near the BZ axis progressively increases moving from the $\Gamma$-point toward $(\pi,0)$ or $(0,\pi)$.  While there is significantly less intensity 
below the QPB compared with the hole-doped system, one can also infer changes to the HEA momentum space position and its ``waterfall"-like appearance 
from cuts (a)-(e) that would be qualitatively similar to the hole-doped system.  The most noticeable contrast to the hole-doped system is the change 
in the HEA energy scale between cuts (a) and (e), a factor on the order of $2$ or $3$, seen directly in the cut along the BZ axis (cut (a) of 
Fig.~\ref{fig:figure5}).  As in the hole-doped case, these results qualitatively capture the behavior observed in experiments on electron-doped 
compounds. \cite{Fujimori_anomaly}

The results presented in this study provide evidence for an HEA in the single-band Hubbard model.  The findings appear to agree qualitatively  
(quantitatively with a proper choice for the energy scale $t$) with experiments on both hole- and electron-doped compounds as well as the half-filled 
parent insulators.  In principle, the results would include effects associated with the coupling of electrons to spin fluctuations, one of the 
theoretical scenarios for the origin of the HEA, at an energy scale proportional to the superexchange constant $J$ that should have a similar value in 
hole- and electron-systems.  However, the dichotomy in the HEA energy scale between hole- and electron-doped systems (a factor of $\sim2$) would argue 
against solely this origin due to bosonic-mode coupling.  Instead these results support the conclusion that strong correlations play a central 
role in the origin of the HEA and that the anomaly itself results from a simple cross-over between the shallow, dispersing QPB near the Fermi level and 
the incoherent LHB at higher binding energy.

The authors would like to thank R.~Hackl, M.~Jarrell, C.~Kim, W.~S.~Lee, A.~Macridin, T.~Maier, W.~Meevasana, R.~T.~Scalettar, F.~Schmitt, Z.-X. 
Shen, and F.~Vernay for valuable discussions. This work is supported in part by the U.S. Department of Energy, Office of Basic Energy Sciences, 
Division of Materials Sciences and Engineering, under contract DOE DE-AC02-76SF00515. SJ would like to acknowledge support from NSERC and SHARCNET.  
This work was made possible, in part, by computational support from the National Energy Research Scientific Computing Center, which is supported by 
the Office of Science of the U.S. Department of Energy under Contract No. DE-AC02-05CH11231,  NSF through TeraGrid resources provided by the National 
Center for Supercomputing Applications, R\'{e}seau Qu\'{e}b\'{e}cois de Calcul de Haute Performance through Mammouth-parall\`{e}le at the University 
of Sherbrooke and the facilities of SHARCNET. The authors wish to thank the Walther Mei{\ss}ner Institute and the Pacific Institute of Theoretical 
Physics for their hospitality during part of this work.

\bibliographystyle{elsarticle-num}
\bibliography{Moritz_HEA}

\end{document}